\documentstyle[12pt,worldsci,epsfig]{article}
\begin{document}
\title{FIRST PRINCIPLE $ q\bar{q} $ SALPETER EQUATION,
SPECTRUM AND REGGE TRAJECTORIES.}
\author{\underline{M. Baldicchi} and G.M. Prosperi\\
{\em Dipartimento di Fisica dell'Universit\`{a}
di Milano and I.N.F.N.\\
Via Celoria,16 20133 Milano, Italy}}
\maketitle
\setlength{\baselineskip}{2.6ex}

\vspace{0.7cm}
\begin{abstract}
Starting from a first principle Salpeter equation we computed
the heavy-heavy, light-light and heavy-light quarkonium spectrum
and the ground light-light Regge trajectories.
We neglect spin-orbit structures and exclude from our
treatment the light pseudoscalar states which in principle
would require the use of the full Bethe-Salpeter
equation due to the chiral symmetry breaking problem.
We found a good agreement with the experimental meson masses
and straight Regge trajectories with the right slope
and intercepts.
Since the parameter are completely specified by the heavy
quarkonium spectrum and by high energy scattering,
for what concerns light-light and light-heavy quarkonium
systems our calculation is essentially parameter free.
\end{abstract}
\vspace{0.7cm}
%
%
%
In preceding papers \cite{prosp96} $ q \bar{q} $
Bethe-Salpeter and $ q $ Schwinger-Dyson equations
were obtained essentially from first principle in a
Wilson loop context.
Only the assumption $  i \ln \! W = i ( \ln \! W )_{ \rm{pert} }
+ \sigma S_{ \rm{min} } $ on the evaluation of the
logarithm of the Wilson loop correlator
as the sum of a perturbative and an area term was used.
Unfortunately a direct treatment of the full B-S equation in
its four dimensional form seems to be beyond the present
possibilities.
From it, however, by a standard method a Salpeter equation was
derived with the Hamiltonian of the form
\begin{equation}
  H = \sqrt{ m_{1}^{2} + {\bf k}^{2} } +
  \sqrt{ m_{2}^{2} + {\bf k}^{2} } + V,
\label{eq1a}
\end{equation}
with
\begin{eqnarray}
&&  \big< {\bf k} | V | {\bf k'} \big> =
  \frac{1}{2 \sqrt{ w_{1} w_{2}
  w_{1}^{\prime} w_{2}^{\prime} } }
  \bigg\{ \frac{4}{3} \frac{ \alpha_{ \rm{s} } }{\pi^{2}}
  \left[ - \frac{1}{ {\bf Q}^{2} } \left( q_{10}
  q_{20} + {\bf q}^{2} -
  \frac{({\bf Q} \cdot {\bf q})^{2}}{ {\bf Q}^{2} }
  \right) + \right.
\nonumber  \\
&&  + \frac{i}{2 {\bf Q}^{2} } {\bf k'} \times {\bf k} \cdot (
  \mbox{\boldmath $ \sigma $}_{1}
  +
  \mbox{\boldmath $ \sigma $}_{2}
  ) + \frac{1}{2 {\bf Q}^{2}} \bigg[ q_{20} (
  \mbox{\boldmath $ \alpha $}_{1}
  \cdot {\bf Q}) - q_{10} (
  \mbox{\boldmath $ \alpha $}_{2}
  \cdot {\bf Q} ) \bigg] +
\nonumber   \\
&&  +  \frac{1}{6} \left.
  \mbox{\boldmath $ \sigma $}_{1}
  \cdot
  \mbox{\boldmath $ \sigma $}_{2}
  + \frac{1}{4}
  \left( \frac{1}{3}
  \mbox{\boldmath $ \sigma $}_{1}
  \cdot
  \mbox{\boldmath $ \sigma $}_{2} -
  \frac{( {\bf Q} \cdot
  \mbox{\boldmath $ \sigma $}_{1}
  ) ( {\bf Q} \cdot
  \mbox{\boldmath $ \sigma $}_{2}
  )}{ {\bf Q}^{2} }
  \right) +
  \frac{1}{4 {\bf Q}^{2} } (
  \mbox{\boldmath $ \alpha $}_{1}
  \cdot {\bf Q}) (
  \mbox{\boldmath $ \alpha $}_{2}
  \cdot {\bf Q} ) \right] +
\label{due} \\
&&  +  \frac{1}{( 2 \pi )^{3}}
  \int \! d^{3} {\bf r} \, e^{i {\bf Q} \cdot {\bf r} }
  J^{ \rm{inst} }({\bf r},{\bf q},q_{10},q_{20}) \bigg\}
\nonumber
\end{eqnarray}
and
\begin{eqnarray}
&&  J^{ \rm{inst} }({\bf r},{\bf q},q_{10},q_{20}) =
  \frac{ \sigma r }{ q_{10} + q_{20} }
  \bigg[ q_{20}^{2} \sqrt{ q_{10}^{2} - {\bf q}_{ \rm{T} }^{2} } +
  q_{10}^{2} \sqrt{ q_{20}^{2} - {\bf q}_{ \rm{T} }^{2} } +
\nonumber \\
&&  + \frac{ q_{10}^{2} q_{20}^{2} }{ | {\bf q}_{ \rm{T} } | }
  \bigg( \arcsin \frac{ | {\bf q}_{ \rm{T} } | }{q_{10}} +
  \arcsin \frac{ | {\bf q}_{ \rm{T} } | }{q_{20}} \bigg) \bigg]
  - \frac{ \sigma }{ r } \bigg[
  \frac{ q_{20} }{ \sqrt{ q_{10}^{2} - {\bf q}_{ \rm{T} }^{2} } }
  \bigg( {\bf r} \times {\bf q} \cdot
  \mbox{\boldmath $ \sigma $}_{1}
  + i q_{10} ( {\bf r} \cdot
  \mbox{\boldmath $ \alpha $}_{1}
  ) \bigg) +
\nonumber \\
&&  + \frac{ q_{10} }{ \sqrt{ q_{20}^{2} - {\bf q}_{ \rm{T} }^{2} } }
  \bigg( {\bf r} \times {\bf q} \cdot
  \mbox{\boldmath $ \sigma $}_{2}
  - i q_{20} ( {\bf r} \cdot
  \mbox{\boldmath $ \alpha $}_{2}
  ) \bigg) \bigg] .
\label{tre}
\end{eqnarray}

In eq.(\ref{eq1a}-\ref{tre})
the perturbative term has been evaluated only at the first order
in the coupling constant $ \alpha_{\rm s} $,
the indices 1 and 2 denote the quark and the antiquark,
$ {\bf k'} $ and $ {\bf k} $ denote
the final and the initial center of mass
momentum of the quark,
$
 w_{j} = \sqrt{ m_{j}^{2} + {\bf k}^{2} },
 w_{j}^{\prime} = \sqrt{ m_{j}^{2} + {\bf k}^{\prime 2} },
 j = 1,2
$.
$ \sigma $ is the string tension,
$
  {\bf q} = \frac{ {\bf k} + {\bf k'} }{2}
$;
$
  {\bf Q} = {\bf k'} - {\bf k}
$;
$
  q_{j0} = \frac{ w_{j} + w_{j}^{\prime} }{2},
$
$
  q_{ \rm{T} }^{h} = ( \delta^{h k} - \hat{r}^{h} \hat{r}^{k} ) q^{k}
$
is the transverse momentum, while
$ \alpha_{j}^{k} $ are the usual Dirac matrices $ \gamma_{j}^{0}
\gamma_{j}^{k} $, and $ \sigma_{j}^{k} $ the $ 4 \times 4 $ Dirac spin
matrices
$ i/4 \, \varepsilon^{knm} [ \gamma^{n}_{j} , \gamma^{m}_{j} ] $.
In the heavy masses limit this potential reduces to that discussed
in ref. \cite{prspcomo1} that is known to reproduce well the spin
averaged multiplets \cite{tstrlt} in the heavy quarkonium cases
and gives the correct fine and
hyperfine splittings \cite{olss93} if even the second order
contributions in $ \alpha_{\rm s} $ are included.
On the other hand if we consider only the confinement part of the
Hamiltonian (\ref{eq1a}-\ref{tre}) in the $ m_{1} = m_{2} = 0 $
limit, setting $ {\bf k} = {\bf k}^{\prime} $ in the cinematic
factors, we obtain
$
  H_{ \rm{hqt} } = 2 | {\bf q} | + \frac{ \pi }{4} \sigma r
$
that gives straight Regge trajectories with slope
\begin{equation}
  \alpha^{\prime} = \frac{ 1 }{ 8 \frac{ \pi }{ 4 } \sigma }
   = \frac{ 1 }{ 2 \pi \sigma }
\label{pend}
\end{equation}
instead of the slope $ \alpha^{\prime} = 1/8 \sigma $ given by
the more commonly used linear rising potential $ V = \sigma r $.
The slope (\ref{pend}) is the same obtained in the
Nambu-Goto string model and gives the experimental slope
$ \alpha^{\prime} = 0.88 $ GeV$ ^{- 2} $ for
$ \sigma \simeq 0.18 $ GeV$ ^{2} $, which is the value used
e.g. in ref \cite{tstrlt}.
All these observations make us hope that the Hamiltonian
(\ref{eq1a}-\ref{tre}) will reproduce well both the
heavy-heavy, light-light and heavy-light quarkonium spectrum
and the light-light Regge trajectories.
We have shown it with a direct numerical computation.
Our results have been reported in ref \cite{mio}, where some
details on the way followed to overcome the numerical difficulties
were explained \cite{lucha96,maung}.

We have adopted the following parameters:
$ \alpha_{\rm s} = 0.363 $,
$ \sigma = 0.175 $ GeV$^{2} $,
$ m_{c} = 1.405 $ GeV,
$ m_{b} = 4.81 $ GeV,
$ m_{s} = 200 $ MeV,
$ m_{u} = 10 $ MeV;
no {\sl ad hoc} constant $ C $ has been added to the potential.
The first four values have to be compared with those
obtained from heavy quarkonium fits (e.g. \cite{tstrlt,tubolsson}).
The light quark masses are taken as the average values of the
current masses reported from the Particle Data Group \cite{prtdatb}.
No attempt of optimizing the parameter was made.

\begin{figure}
\begin{centering}
    \leavevmode
    \setlength{\unitlength}{1.0mm}
    \begin{picture}(140,70)
      \put(25,0){\mbox{\epsfig{file=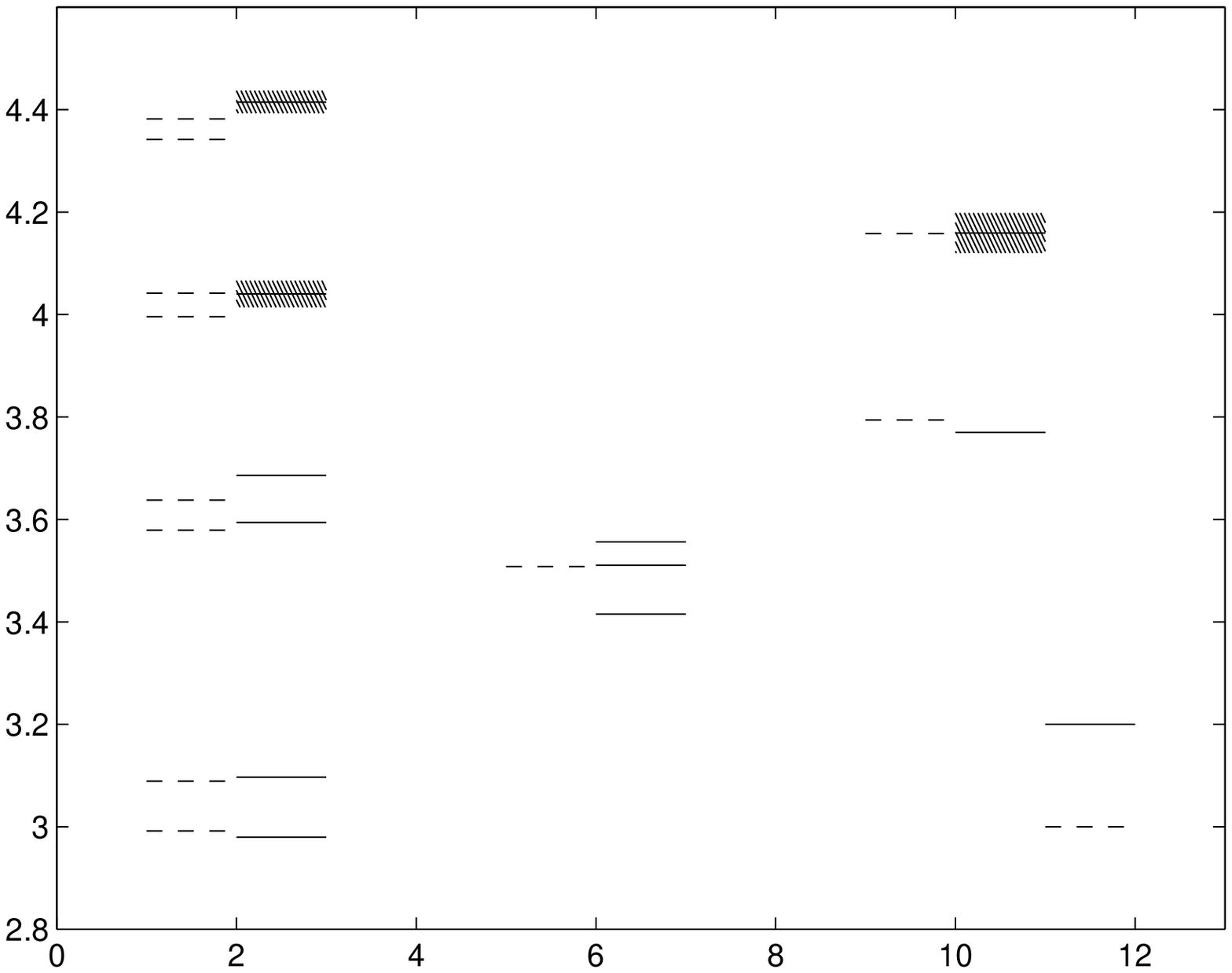,height=7cm}}}
      \put(16,65){GeV}
      \put(16,55){M}
      \put(70,61){ \Large{$ c \bar{c} $ } }
      \put(88,10){theor.}
      \put(88,17){exper.}
      \put(48,8){ $ 1 \, {^{1} {\rm S}_{0}} $ }
      \put(48,13){ $ 1 \, {^{3} {\rm S}_{1}} $ }
      \put(48,30){ $ 2 \, {^{1} {\rm S}_{0}} $ }
      \put(48,35){ $ 2 \, {^{3} {\rm S}_{1}} $ }
      \put(48,47){ $ 3 \, {^{3} {\rm S}_{1}} $ }
      \put(48,61){ $ 4 \, {^{3} {\rm S}_{1}} $ }
      \put(74,23){ $ 1 \, {^{3} {\rm P}_{0}} $ }
      \put(74,27){ $ 1 \, {^{3} {\rm P}_{1}} $ }
      \put(74,31){ $ 1 \, {^{3} {\rm P}_{2}} $ }
      \put(100,38){ $ 1 \, {^{3} {\rm D}_{1}} $ }
      \put(100,52){ $ 2 \, {^{3} {\rm D}_{1}} $ }
    \end{picture}
\caption{ $ c \bar{c} $ quarkonium spectrum. }
\label{fig4}
\end{centering}
\end{figure}

\begin{figure}
\begin{centering}
    \leavevmode
    \setlength{\unitlength}{1.0mm}
    \begin{picture}(140,70)
      \put(25,0){\mbox{\epsfig{file=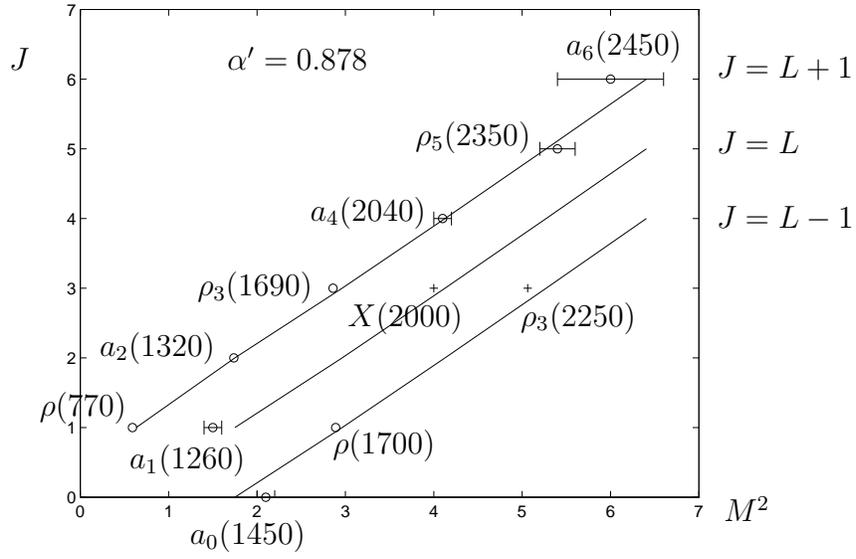,height=7cm}}}
      \put(16,60){ $ J $ }
      \put(110,59){ $ J = L + 1 $ }
      \put(110,49){ $ J = L $ }
      \put(110,39){ $ J = L - 1 $ }
      \put(45,60){ $ \alpha^{ \prime } = 0.878 $ }
      \put(111,0){ $ M^{2} $ }
      \put(20,14){ $ \rho ( 770 ) $ }
      \put(28,22){ $ a_{2} ( 1320 ) $ }
      \put(32,7){ $ a_{1} ( 1260 ) $ }
      \put(40,-3){ $ a_{0} ( 1450 ) $ }
      \put(41,30){ $ \rho_{3} ( 1690 ) $ }
      \put(59,9){ $ \rho ( 1700 ) $ }
      \put(56,40){ $ a_{4} ( 2040 ) $ }
      \put(61,26){ $ X ( 2000 ) $ }
      \put(70,50){ $ \rho_{5} ( 2350 ) $ }
      \put(84,26){ $ \rho_{3} ( 2250 ) $ }
      \put(90,62){ $ a_{6} ( 2450 ) $ }
    \end{picture}
\caption{Ground triplet $ u \bar{u} $ Regge trajectories.
Theoretical results (full line) compared with experimental
data (circlet). Cross denote less established masses.}
\label{fig1}
\end{centering}
\end{figure}

As an example for the heavy-heavy systems we report in fig.
\ref{fig4} the $ c \bar{c} $ spectrum and
for the light-light systems we report in fig.
\ref{fig1} the $ u \bar{u} $ Regge trajectories.
However we obtained similar good results
also for the other treated systems \cite{mio}.
The results for heavy-light systems
are reported in table \ref{tablegn} and compared with the
experimental spin averaged masses using the theoretical
splitting collected in table \ref{tableg8}
when the singlet state has not yet been observed.
%
%
%
%
\begin{table}
\centering
\caption{Theoretical results for 
$ q \bar{q} $ hyperfine splitting (MeV).
Experimental data are enclosed in brackets.}
\begin{tabular}{ccccccc}
\hline
 State & $ u \bar{c} $ & $ u \bar{b} $ & $ c \bar{c} $ &
 $ b \bar{b} $ & $ s \bar{c} $ & $ s \bar{b} $ \\
\hline
 1S &
 111 ($ 141 \pm 1 $) &
 59 ($ 46 \pm 3 $) &
 97 ($ 117 \pm 2 $) &
 102 &
 108 (144) &
 60 ($ 47 \pm 4) $    \\
 2S &
 59 &
 38 &
 59 ($ 92 \pm 5 $) &
 42 &
 62 &
 40 \\
\hline
\end{tabular}
\label{tableg8}
\end{table}
\begin{table}
\caption{Theoretical results for
$ u \bar{c} $, $ u \bar{b} $, $ s \bar{c} $, $ s \bar{b} $ systems
(MeV). Experimental data are enclosed in brackets.}
\begin{tabular}{ccccc}
\hline
 State & $ u \bar{c} $ & $ u \bar{b} $ &
 $ s \bar{c} $ & $ s \bar{b} $ \\
\hline
 1S &
 1973 ($ 1973 \pm 1 $) &
 5326 ($ 5313 \pm 2 $) &
 2080 ($ 2076.4 \pm 0.5 $) &
 5418 ($ 5404.6 \pm 2.5) $
\\
 2S &
 2600 $ ( 2623 \pm ? )^{\rm a} $ &
 5906 $ ( 5897 \pm ? )^{\rm a} $ &
 2713 &
 6004
\\
 1P &
 2442 $ ( 2438 \pm ? )^{\rm b} $ &
 5777 $ ( 5825 \pm 14 )^{\rm c} $ &
 2528 ($ 2535.35 \pm 0.34 $)  &
 5848 ($ 5853 \pm 15 $)
\\
\hline
\end{tabular}
\label{tablegn}
$ ^{\rm a} $Obtained from preliminary {\sl Delphi} data
$ m(D^{\ast \prime}) = 2637 \pm 8 $ MeV,
$ m(B^{\ast \prime}) = 5906 \pm 14 $ MeV \cite{pullia}
subtracting 1/4 theoretical hyperfine splitting reported in
table \ref{tableg8}.
\\
$ ^{\rm b} $Estimated from
$ m(D_{2}^{\ast}) = 2459 \pm 4 $ MeV,
$ m(D_{1}) = 2427 \pm 5 $ MeV.
\\
$ ^{\rm c} $From preliminary {\sl Delphi} data \cite{pullia}.
\end{table}

In conclusion, starting from a first principle Salpeter
equation, we have obtained a good reproduction of the
meson spectrum involving heavy and light quarks with
the exception of the light pseudoscalar states.
Also the light-light Regge trajectories are well reproduced.
Since the parameter are completely specified by the heavy
quarkonium spectrum and by high energy scattering,
for what concerns light-light and light-heavy quark systems
our calculation is essentially parameter free.
%
%
%
%
\end{document}